\documentclass[12pt]{iopart}

\usepackage{iopams}
\usepackage{graphicx}
\begin{document}

\title[]{Exact solution of a delay difference equation modeling traffic flow and their ultra-discrete limit}

\author{Keisuke Matsuya$^1$ and Masahiro Kanai$^2$}
\address{$^1$Faculty of Engineering, Department of Mathematical Engineering, Musashino University, 3-3-3 Ariake, Koto-ku Tokyo 135-8181, Japan}
\address{$^2$Graduate School of Mathematical Sciences, The University of Tokyo, 3-8-1 Komaba, Tokyo 153-8914, Japan}
\ead{$^1$keimatsu@musashino-u.ac.jp}
\ead{$^2$kanai@ms.u-tokyo.ac.jp}


\begin{abstract}
We consider a car-following model described by a delay difference equation and give its exact solutions that present propagation of a traffic jam.
This model is a discrete-time version of the delayed optimal-velocity model; in the continuum limit, we recover the delay differential equation for this model and the exact solutions as well.
We then work in the ultra-discrete limit, obtaining a delay cellular-automaton model, which successfully inherits the solutions.
Also the dispersion relation for the present solutions suggests that a quick response of drivers does not always result in fast dissolution of a traffic jam.
\end{abstract}

%
\vspace{2pc}
%
%
%
%

\section{Introduction}

Study of traffic flow relies in no small part on theoretical/mathematical modeling because of the difficulty in extensive measurements and designed experiments, and predictions from simulation and analytical study have particular importance in this field \cite{SCN}.

A car-following model is described in general by an equation of the form:
\[ \dot{x}_n(t)=G\left( h_n(t-\tau) \right) \qquad ( h_n(t):= x_{n+1}(t)-x_n(t) ). \]
$x_n(t)$ denotes the position of the $n$th car at time $t$ and a positive constant $\tau$ is a delay in time, which corresponds to the driver's reaction time.
After this latency, each car reaches the velocity $G( h_n(t-\tau) )$ determined by the headway $h_n(t-\tau)$, i.e., a distance to the next car in front;
 hence $G(\cdot)$ is nowadays referred to as {\it the optimal-velocity (OV) function}.

In the earlier studies of traffic flow, G. F. Newell proposed the car-following model with
\[ G(h)= V_0\left( 1-\exp\left( -\frac{\gamma}{V_0} (h-L) \right) \right), \]
where $V_0$, $\gamma$ and $L$ are positive constants, and found an exact solution for $\tau=0$ \cite{Newell}.
In this case, the equation can be transformed into a linear equation and thus allows for superposition of the shocks each of which present a traffic jam propagating upstream at different speeds.

Subsequently, G. B. Whitham obtained an elliptic solution for Newell's model in the generic case $\tau\ne0$ \cite{Whitham}.
This presents nonlinear waves beyond Newell's {\it linear} solutions, including a solitary-wave in the case that the modulus of the elliptic function goes to 1.
He also suggested that the solitary-wave solution be stable in the range of parameters for stable linear waves.
Both Newell's model and the Kac--van Moerbeke system reduce to the same equation if one assumes a steady-profile solution with the phase velocity $1/2\tau$; the Kac--van Moerbeke system is known as one of the integrable lattice systems \cite{KvM}.
This fact motivated Whitham to derive a solitary-wave solution of Newell's model after a long calculation.
(It is notable that these results consist with the theory of nonlinear waves \cite{NonlinearWaves}.)

Bando {\it et al.} proposed an alternative car-following model described by a second-order differential equation $\ddot{x}_n(t)=a[V(h_n(t))-\dot{x}_n(t)]$ with the sensitivity $a$ (a positive constant) and the OV function,
\begin{equation}\label{OVF}
V(h)=\tanh(h-c)+\tanh c,
\end{equation}
where $c>0$ is the inflection point of the {\it tanh} function.
They referred to the model as the optimal-velocity (OV) model \cite{Bando1,Bando2}.
(Note that replacing $\dot{x}_n(t+\tau)$ with $\ddot{x}_n+a x_n$ for small $\tau=1/a$, one formally recovers the OV model.)
Now $V(h)$ is the most popular choice for the OV function and is actually suitable for mathematical treatment.
In \cite{Hasebe,Nakanishi}, they showed exact solutions for the car-following model with the OV function \eref{OVF},
\[ \dot{x}_n(t)=V(h_n(t-\tau)), \]
which they call {\it the delayed optimal-velocity (OV) model}.

The car-following model is totally asymmetric in the lattice variable, and which is one of the crucial properties of traffic flow models.
However, in the above works, a steady-profile solution is chosen so that the delay in time should compensate for the asymmetric lattice shift.
Consequently, all the solutions obtained, given by elliptic functions, have the same phase velocity of $1/2\tau$ independently of the form of the OV function.
An alternative way to exactly solve a traffic-flow model is to assume a shock-wave solution.

Inspired by Hirota's method \cite{Hirota} to find the soliton solutions of soliton equations, in \cite{Tutiya} they reduced the delayed OV model expressed in $h_n$,
\begin{equation}\label{dOV}
\eqalign{
\dot{h}_n(t+\tau)&=V(h_{n+1}(t-\tau))-V(h_n(t-\tau))\\
&=\tanh(h_{n+1}(t-\tau)-c)-\tanh(h_n(t-\tau)-c), 
}
\end{equation}
to two coupled equations, and thus found a shock-wave solution.
It is distinct from the previous results introduced above that their solution includes a free parameter which can control the shock velocity.
They also considered the same scenario for Newell's model, finding a shock solution with velocity $1/\tau$; this solution also includes a free parameter but does not allow for the variation of the velocity.

In this work, we solve exactly the discrete delayed OV model firstly introduced in \cite{UDOV} and thereby integrate the previous results of \cite{Tutiya} and \cite{UDOV}.
In Section 2, we derive the discrete delayed OV model and give exact results.
We also consider the continuum limit as the discrete delayed OV model reduces to the delayed OV model.
In Section 3, we consider the ultra-discrete limit of the results obtained in the previous section.
In this limit, the model is transformed into a cellular automaton; thus we obtain a delay cellular automaton with an exact solution.
Section 4 is devoted to summary and conclusion.

\section{Discretization of the delayed OV model and its exact solution}

Changing the variable as 
\begin{equation}\label{tanh}
g_n(t) = \tanh{(h_n(t)-c)},
\end{equation}
 in \eref{dOV},  we transform the delayed OV model into an algebraic differential equation:
\begin{equation}\label{dOV2}
\dot{g}_n(t) = \left(1-g_n(t)^2\right)\left(g_{n+1}(t-\tau)-g_n(t-\tau)\right),
\end{equation}
where the transcendental function {\it tanh} disappears.
Then we discretize \eref{dOV2} and obtain its exact solution assuming the open boundary condition that an infinite number of cars run on the road, i.e., $n\in\mathbb{Z}$.

\subsection{Discretization of the delayed OV model}

First we shall review the relationship between \eref{dOV2} and {\it the modified Lotka--Volterra (mLV) equation} given by
\begin{equation}\label{mLV}
\dot{r}_j(t) = r_j(t)\left(1-ar_j(t)\right)\left(r_{j+1}(t)-r_{j-1}(t)\right),
\end{equation}
where $j\in\mathbb{Z}$ and $a \in \mathbb{R}$.

If one assumes a traveling-wave solution to \eref{dOV2}, i.e., $g_n(t) = G(\phi)$ with $\phi=t+2n\tau$, \eref{dOV2} yields
\begin{equation}\label{dOVtravel}
G^\prime(\phi) = \left(1-G(\phi)^2\right)\left(G(\phi+\tau)-G(\phi-\tau)\right),
\end{equation}
where ${}^\prime$ stands for the derivative.
Meanwhile, changing the variable as $r_j(t)=(1+\bar{r}_j(t))/(2a)$, from \eref{mLV} we have 
\begin{equation}\label{mLV2}
4a\dot{\bar{r}}_j(t) = \left(1-\bar{r}^2_j(t)\right)\left(\bar{r}_{j+1}(t)-\bar{r}_{j-1}(t)\right).
\end{equation}
Then, we assume the traveling-wave solution, $\bar{r}_j=\bar{R}(\psi)$ with $\psi:=t/(4a)+j\tau$, to \eref{mLV2} and reach the same equation as \eref{dOVtravel}:
\[\bar{R}'(\psi)=\left(1-\bar{R}(\psi)^2\right)\left(\bar{R}(\psi+\tau)-\bar{R}(\psi-\tau)\right).\]
We thus see that both \eref{dOV2} and \eref{mLV} reduce to the same equation \eref{dOVtravel} if a traveling-wave solution is assumed.

Here we introduce {\it the discrete modified Korteweg--de Vries (mKdV) equation} \cite{Tsujimoto},
\begin{equation}\label{dmKdV}
v^{t+1}_j\frac{1+\delta v^{t+1}_{j+1}}{1-av^{t+1}_j} = v^t_j\frac{1+\delta v^t_{j-1}}{1-av^t_j},
\end{equation}
where $\delta\in\mathbb{R}$.
This is known as a discretization of the mLV equation \eref{mLV}.
(We use superscript $t$ as the discrete time variable.)
Actually, if we let $v^t_j=r_j(-\delta t)$ then \eref{dmKdV} reduces to \eref{mLV} in the continuum limit where $\delta$ regarded as the discrete time unit tends to 0.

In the same manner as above, we change the variable as $v^t_j=(1+\bar{v}^t_j)/(2a)$ in \eref{dmKdV} and have
\begin{equation}\label{dmKdV2}
\frac{1-2\gamma}{\gamma} (\bar{v}^{t+1}_j-\bar{v}^t_j) = (1-\bar{v}^t_j)(1+\bar{v}^{t+1}_j)\bar{v}^{t+1}_{j+1} - (1-\bar{v}^{t+1}_j)(1+\bar{v}^t_j)\bar{v}^t_{j-1}.
\end{equation}
where $\gamma:=\delta/(4a)$.
Assume the traveling-wave solution, $\bar{v}^t_j=\bar{V}(\Psi)$ with $\Psi:=t+mj$ ($m\in\mathbb{Z}_{>0}$), and \eref{dmKdV2} yields
\begin{equation}\label{dmKdVtravel}
\fl\eqalign{
\frac{1-2\gamma}{\gamma}\left(\bar{V}(\Psi+1)-\bar{V}(\Psi)\right) = \left(1-\bar{V}(\Psi)\right)\left(1+\bar{V}(\Psi+1)\right)\bar{V}(\Psi+1+m) \cr
\qquad\qquad\qquad\qquad\qquad\qquad\qquad\qquad- \left(1-\bar{V}(\Psi+1)\right)\left(1+\bar{V}(\Psi)\right)\bar{V}(\Psi-m).
}
\end{equation}
Here we introduce an alternative variable $u^t_n$ and then consider $u^t_n=U(\Phi)$ with $\Phi:=t+2mn$.
If $U(\Phi)$ satisfies \eref{dmKdVtravel}, we have
\begin{equation}\label{fdOVtravel}
\fl\eqalign{
\frac{1-2\gamma}{\gamma}\left(U(\Phi+1)-U(\Phi)\right) = \left(1-U(\Phi)\right)\left(1+U(\Phi+1)\right)U(\Phi+1+m) \cr
\qquad\qquad\qquad\qquad\qquad\qquad\qquad\qquad- \left(1-U(\Phi+1)\right)\left(1+U(\Phi)\right)U(\Phi-m).
}
\end{equation}
Since $u^t_n=U(\Phi)$ ($\Phi=t+2mn$), we may replace $U(\Phi+1+m)$ with $u^{t+1-m}_{n+1}$ and $U(\Phi-m)$ with $u^{t-m}_n$, and consequently obtain {\it the discrete delayed optimal-velocity (OV) model}:
\begin{equation}\label{fdOV}
\fl\frac{1-2\gamma}{\gamma}\left(u^{t+1}_n-u^t_n\right) = \left(1-u^t_n\right)\left(1+u^{t+1}_n\right)u^{t-m+1}_{n+1} - \left(1-u^{t+1}_n\right)\left(1+u^t_n\right)u^{t-m}_n.
\end{equation}
Note that \eref{fdOV} possesses the asymmetry in the lattice variable $n$ which is required for the car-following model.

Let us see that the discrete delayed OV model \eref{fdOV} is the discrete version of the delayed OV model \eref{dOV2}.
We restrict ourselves to $\gamma>0$, for $\gamma$ corresponds to the discrete time unit.
If there exists a smooth function $\tilde{g}_n(t)$ that satisfies $\tilde{g}_n(\gamma t)=u^t_n$, we have
\begin{equation}\label{c.l.}
\fl\eqalign{
\frac{1-2\gamma}{\gamma}\left(\tilde{g}_n(t+\gamma)-\tilde{g}_n(t)\right) = \left(1-\tilde{g}_n(t)\right)\left(1+\tilde{g}_n(t+\gamma)\right)\tilde{g}_{n+1}(t-\tau+\gamma) \cr
\qquad\qquad\qquad\qquad\qquad\qquad\qquad\qquad- \left(1-\tilde{g}_n(t+\gamma)\right)\left(1+\tilde{g}_n(t)\right)\tilde{g}_n(t-\tau),
}
\end{equation}
where we let $\tau:= m\gamma$.
Consequently, taking the limit $\gamma \to +0$, we successfully recover \eref{dOV2} from \eref{c.l.}.

Accordingly, the discrete variable $h^t_n$ corresponding to the headway $h_n(t)$ should be defined from \eref{tanh} as
\begin{equation}\label{dheadway}
h^t_n=c+\frac12\log{\frac{1+u^t_n}{1-u^t_n}}.
\end{equation}

\subsection{Exact solution of the discrete delayed OV model}

In order to find shock solutions of \eref{fdOV}, we assume 
\begin{equation}\label{sol}
u^t_n = \frac{M + N L^tK^n}{1 + L^tK^n},
\end{equation}
where $K\neq1$, $L\neq1$, $M$ and $N$ ($M \neq N$) are constants to be determined.
Since \eref{sol} can transform into $\frac{N + M (L^{-1})^t(K^{-1})^n}{1 +  (L^{-1})^t(K^{-1})^n}$, we only consider the case of $L>1$.
Substituting \eref{sol} into \eref{fdOV}, we obtain
\begin{equation}\label{relation1}
\fl\eqalign{
(LK-1)M^2-2L^m(L-1)M+L^m(L-1)\Delta-(LK-1)=0,\cr
L^m(LK-1)N^2-2(L-1)KN+(L-1)K\Delta-L^m(LK-1)=0,\cr
(L+1)(LK-1)MN-(LK+1)(L-1)(M+N) \cr
\qquad\qquad\qquad\qquad~+(LK+1)(L-1)\Delta-(L+1)(LK-1)=0,
}
\end{equation}
where $\Delta:=(1-2\gamma)/\gamma$.
Elimination of $M$ and $N$ from \eref{relation1} yields the two cases:
\begin{eqnarray}
K=L^{-2},\label{relation2}\\
K=\frac{L-1-4\gamma (L^{m+1}-1)}{L( L-1-4\gamma(L- L^{-m}) )}.
\label{relation3}
\end{eqnarray}
These are the dispersion relations for the traveling-wave solutions \eref{sol}.

Remember that $u^t_n$ corresponds to \eref{tanh} and that the distance $h_n(t)$ of cars must be positive, and accordingly we shall find the solution in the range of
\begin{equation}\label{range}
-\tanh c< u^t_n< 1.
\end{equation}
In the case of \eref{relation2}, from \eref{relation1} we obtain two shock solutions; however these lie out of the range \eref{range}.

\begin{figure}[ht]
\includegraphics[width=\textwidth]{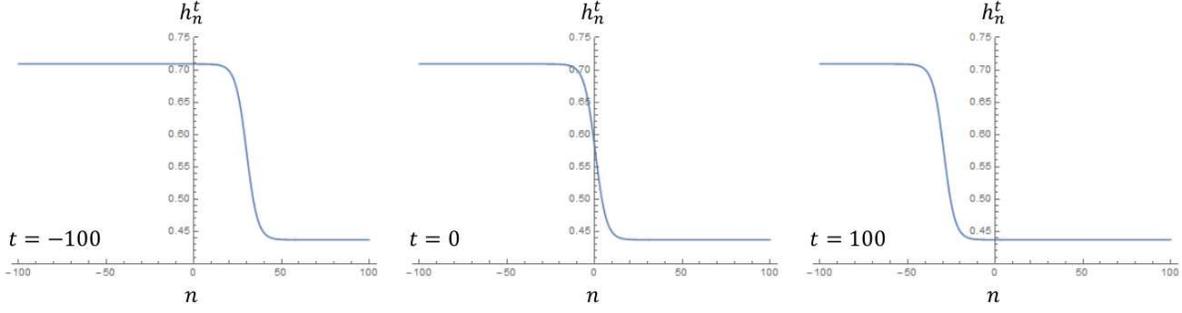}
\caption{Exact shock solution for the discrete delayed OV model: the headway $h^t_n$ \eref{dheadway} of $n$th car where $u^t_n$ is given by \eref{sol2} with parameters $c=1$, $L=1.1$, $\gamma=0.2$, $m=3$ at time $t=-100$, $t=0$, $t=100$.}
\label{figdiscrete}
\end{figure}

In the case of \eref{relation3}, together with $M$ and $N$ determined by \eref{relation1} we obtain the two shock solutions:
\begin{eqnarray}
u^t_n=1-\frac{(1-4\gamma)(L-1)}{2\gamma(1-L^{-m})} \frac{1+K^nL^{t-m-1}}{1+K^nL^t},
\label{sol1}\\
u^t_n=-1+ \frac{L-1}{2\gamma (L-L^{-m})} \frac{1+ K^n L^{t-m}}{1+K^n L^t}.
\label{sol2}
\end{eqnarray}
Figure \ref{figdiscrete} shows the graphs of an exact solution for the discrete delayed OV model: the headway $h^t_n$ given by \eref{dheadway} and \eref{sol2}.
This illustrates a shock wave propagating upstream; the shock front corresponds to the tail of a traffic jam.
Alternatively, \eref{sol1} will show an upside-down shock propagating upstream, which illustrates dissolution of a traffic jam; the shock front corresponds to the head of a traffic jam.

The parameter values used in the simulation are chosen to satisfy the conditions discussed below.
In general, the range of a shock wave is bounded by the limit values at infinity.
Hence we have only to investigate these values in order to find the range of the solution.

At first, we consider \eref{sol1}.
Since $L>1$ and $m\in\mathbb{Z}_{>0}$, the upper limit of \eref{sol1} is smaller than $1$ if and only if $1-4\gamma>0$ is satisfied.
Now we assume $1-4\gamma>0$.
Since $L>1$,\ $m\in\mathbb{Z}_{>0}$ and $0<\gamma<1/4$, taking the limits as $n\to\pm\infty$, we find that $u^t_n$ satisfies
\begin{equation*}
1-\frac{(1-4\gamma)(L-1)}{2\gamma(1-L^{-m})}<u^t_n<1-\frac{(1-4\gamma)(L-1)}{2\gamma(1-L^{-m})}L^{-m-1}
\end{equation*}
for all $t$ and $n$.
We introduce the function of $L$,
\begin{equation}\label{f}
f(L):=1-\frac{(1-4\gamma)(L-1)}{2\gamma(1-L^{-m})},
\end{equation}
and then have the derivative
\begin{equation*}
f^\prime(L)=-\frac{(1-4\gamma)(L^{m+1}-(m+1)L+m)}{2\gamma L^{m+1} (1-L^{-m})^2}.
\end{equation*}
This leads to $f^\prime(L)\leq0$ for all $L>1$.
Consequently, if and only if $-\tanh c< f(1)$, we find an $L$ that satisfies $-\tanh c<f(L)$ and a shock solution that satisfies \eref{range}.
From $f(1)=1-(1-4\gamma)/(2m\gamma)$, we explicitly show the condition to have an exact shock solution:
\begin{equation}\label{condition1}
\frac1{4+2m(1+\tanh c)}<\gamma<\frac14.
\end{equation}
(We remark that one can find an {\it unphysical} solution, $|u^t_n|>1$, unless \eref{condition1} is considered.)
Assuming \eref{condition1}, we can find the maximum value $L_f$ of $L$ such that $-\tanh c<f(L)$ is satisfied.
Consequently, we have
\begin{equation}
f(L) > f(L_f)=-\tanh c\qquad (1<L<L_f).
\label{Lhat}
\end{equation}

Next we consider the solution \eref{sol2}.
As well as the above discussion, we find the lower limit of \eref{sol2} is larger than $-1$ if and only if $1-4\gamma>0$ is satisfied.
We also assume $1-4\gamma>0$ in this case.
Taking the limits as $n\to\pm\infty$, we find that $u^t_n$ satisfies
\begin{equation*}
-1+ \frac{L^{-m}(L-1)}{2\gamma (L-L^{-m})}<u^t_n<-1+ \frac{L-1}{2\gamma (L-L^{-m})}
\end{equation*}
for all $t$ and $n$.
Now we let
\[g(L):=-1+\frac{L-1}{2\gamma(L^{m+1}-1)},\]
and thus find that \eref{sol2} ranges from $g(L)$ to $g(L^{-1})$.
Then we see the condition to have an exact solution,
\[\frac1{4(m+1)}<\gamma<\frac1{2(m+1)(1-\tanh c)},\]
and that there exists the maximum value $L_g$ of $L$ such that
\begin{equation}
-\tanh c = g(L_g) < g(L) < g(L^{-1}) < 1 \qquad (1<L<L_g).
\label{Lhat2}
\end{equation}
Since $f(L)>-1$ or $g(L^{-1})<1$ is equivalent to $L-1-4\gamma(L-L^{-m})<0$, we find $K>1$ for $L<\max{(L_f,L_g)}$.

The phase velocity of the shock solutions \eref{sol1} and \eref{sol2} is given by
\[U=\frac{\log L}{\log K}=\frac{\log L}{\log \frac{L-1-4\gamma (L^{m+1}-1)}{L( L-1-4\gamma(L- L^{-m}) )}}. \]
From this formula, we can see that the phase velocity may attain the maximum at an arbitrary $m$, but however it appears complicated.
For simplicity, we will discuss this point for the continuum version.

\subsection{Continuum limit of the shock solution}

As described above, $\delta$ or $\gamma$ takes the role of discrete time unit and tends to 0 when we consider the continuum limit.

The dispersion relation \eref{relation3} implies that $L$ converges to $1$ or $1/K$ in the continuum limit $\gamma \to +0$.
However, it is apparent that $1/K$ is never admissible for \eref{sol1}.
We hence let $L = 1+\beta\gamma+\Or(\gamma^2)$, which gives $L^m\to e^{\beta \tau}$ as $\gamma\to+0$.

Also, let $K=e^\alpha$, $m\gamma=\tau$ and $t$ be replaced with $t/\gamma$ to consider the continuum limit.
(Note that the latter $t$ denotes the continuous time variable.)
Thus we have the exact solutions of \eref{dOV2},
\begin{equation}\label{sol3}
g_n(t)=\pm\left(1-\frac{\beta}{2(1-e^{-\beta\tau})} \frac{1+e^{\alpha n +\beta (t-\tau)}}{1+e^{\alpha n+\beta t}}\right),
\end{equation}
and the dispersion relation,
\begin{equation}\label{relation5}
e^\alpha = \frac{\beta-4(e^{\beta\tau}-1)}{\beta-4(1-e^{-\beta\tau})},
\end{equation}
which is common to both solutions.
The solution \eref{sol3} and the dispersion relation \eref{relation5} obtained here coincide with those given in \cite{Tutiya}.
Also, from \eref{condition1} we have 
\begin{equation}\label{condition2}
\tau>\frac1{2(1+\tanh c)}.
\end{equation}
Note that $c>0$ ensures $\tau>1/4$ and consequently that the right hand side of \eref{relation5} is positive.
For small $\tau$, the phase velocity $U={\beta}/{\alpha}$ may be approximated by a power series in $\tau$:
\[U= -\frac{1}{4\tau^2}+\frac1{\tau}+\frac{\beta^2}{48}+O(\tau).\]
This suggests that a quick response of drivers to the traffic situation does not always yield fast propagation of a traffic jam.
We find the maximum velocity
\[U=\frac{\beta}{2\log(\frac{4+\beta}{4-\beta})}\]
 at $\tau= \frac1\beta \log\frac{4+\beta}{4-\beta}$.

\begin{figure}[htb]
\includegraphics[width=\textwidth]{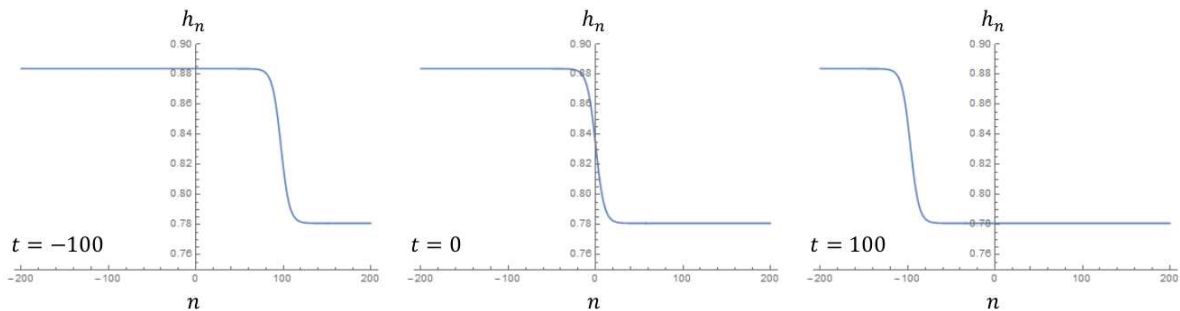}
\caption{Exact shock solution for the delayed OV model: the headway $h_n(t)$ \eref{headway} of $n$th car where $u_n(t)$ is given in \eref{sol3} with minus sign and parameters $c=1$, $\beta=0.2$, $\tau=0.6$ at time $t=-100$, $t=0$, $t=100$.}
\label{figcont}
\end{figure}

Figure \ref{figcont} shows an exact shock solution: the headway
\begin{equation}
h_n(t)=c+\frac12\log\frac{1+g_n(t)}{1-g_n(t)},
\label{headway}
\end{equation}
 which corresponds to the solution with minus sign in \eref{sol3}.
As expected, it illustrates a traffic jam propagating upstream.

\section{Ultra-discrete limit of the discrete delayed OV model}

In this section, we consider the ultra-discrete limit of the discrete delayed optimal-velocity model \eref{fdOV}.
In this limit, the dependent variable is discretized and thus we obtain a delay cellular automaton from the delay discrete equation.

\subsection{The ultra-discrete limit}

The ultra-discrete limit was firstly developed to directly relate the soliton equations to cellular automata \cite{Tokihiro}.

We define {\it the ultra-discrete limit} or {\it ultra-discretization} by the transformation from a positive real variable $u$ to a ultra-discrete variable $U$:
\[U=\lim_{\varepsilon\to+0} \varepsilon\log u,\]
where $\varepsilon$ is the ultra-discrete parameter.
(In usual, a ultra-discrete variable is denoted by the capital letter corresponding to that of the variable.)
By linking $u$ with $U$ by 
\begin{equation}\label{ud}
u=e^{U/\varepsilon},
\end{equation}
multiplication/division and addition in the original variable are transformed respectively into addition and `max' in the ultra-discrete variable:
\[
\lim_{\varepsilon\to+0} \varepsilon\log (u\cdot v)=U+V,\qquad
\lim_{\varepsilon\to+0} \varepsilon\log(u+v)=\max(U,V).
\]
Thus, an equation including multiplication and addition but no subtraction transforms into the ultra-discrete equation described with addition and `max' in the ultra-discrete variable.
Since addition and `max' become integer functions being restricted to integers, we can regard the ultra-discrete equation as a cellular automaton.

\subsection{Ultra-discrete limit of the discrete mLV equation}
As mentioned above, the ultra-discretization runs only for equations described with multiplication and addition.
We hence change the dependent variable as $\tilde{v}^t_j=v^t_j/(1-av^t_j)$ to eliminate subtraction from \eref{dmKdV}:
\begin{equation}\label{dmLV}
\tilde{v}^{t+1}_j\frac{1+(a+\delta)\tilde{v}^{t+1}_{j+1}}{1+a\tilde{v}^{t+1}_{j+1}}
=\tilde{v}^{t}_j\frac{1+(a+\delta)\tilde{v}^{t}_{j-1}}{1+a\tilde{v}^{t}_{j-1}}.
\end{equation}
Then, we introduce the ultra-discrete variable $V^t_j$ and parameters $D$ and $A$ as $\tilde{v}^t_j=e^{V^t_j/\varepsilon}$, $\delta=e^{-D/\varepsilon}$, $a=e^{-A/\varepsilon}$,
and thus obtain from \eref{dmLV} the ultra-discrete mKdV equation,
\[
\eqalign{
&V^{t+1}_{j}+\max(0,V^{t+1}_{j+1}-D,V^{t+1}_{j+1}-A)-\max(0,V^{t+1}_{j+1}-A)\\
&\qquad =V^{t}_{j}+\max(0,V^{t}_{j-1}-D,V^{t}_{j-1}-A)-\max(0,V^{t}_{j-1}-A),
}
\]
where $D$ and $A$ take integer values as well as $V^t_j$ \cite{Takahashi}.

\subsection{Ultra-discrete limit of the discrete delayed OV model}
We remark that the variable changes, $v^t_j=(1+\bar{v}^t_j)/(2a)$, $\tilde{v}^t_j=v^t_j/(1-av^t_j)$ and $\tilde{v}^t_j=e^{V^t_j/\varepsilon}$, done thus far are collected into
\[
\bar{v}^t_j=\frac{a\tilde{v}^t_j-1}{a\tilde{v}^t_j+1}
=\tanh\left(\frac{V^t_j-A}{2\varepsilon}\right).
\]
Taking $g_n(t)=\tanh(h_n(t)-c)$ into account, we introduce the ultra-discrete variable $H^t_n$ corresponding to the car distance $h^t_n$ such that
\[
u^t_n = \tanh{\left(\frac{H^t_n-C}{2\varepsilon}\right)},
\]
where $C>0$ is an ultra-discrete parameter corresponding to $c$, i.e., $h^t_n=H^t_n/(2\varepsilon)$ and $c=C/(2\varepsilon)$.
Then, in order to ultra-discretize \eref{fdOV}, we change the variable as $\tilde{u}^t_n=(1+u^t_n)/(1-u^t_n)$ which means $u^t_n=(\tilde{u}^t_j-1)/(\tilde{u}^t_j+1)$ and
\begin{equation}\label{udl1}
\tilde{u}^t_n = e^{(H^t_n-C)/\varepsilon}.
\end{equation}
Accordingly, from \eref{fdOV} we have
\begin{equation}\label{fdOV2}
\tilde{u}^{t+1}_n \frac{(1-4\gamma)\tilde{u}^{t-m+1}_{n+1}+1}{\tilde{u}^{t-m+1}_{n+1}+1}
 = \tilde{u}^t_n \frac{(1-4\gamma)\tilde{u}^{t-m}_n+1}{\tilde{u}^{t-m}_n+1}.
\end{equation}

The condition \eref{condition1} for there to be an exact shock solution suggests that we should introduce the ultra-discrete parameter $G$ corresponding to $\gamma$ as
\begin{equation}\label{udl2}
1-4\gamma = e^{-G/\varepsilon},
\end{equation}
where $G>0$; this implies that we focus on around $\gamma=1/4$.
Consequently, substitution of \eref{udl1} and \eref{udl2} into \eref{fdOV2}, we obtain {\it the ultra-discrete delayed optimal-velocity (OV) model},
\begin{equation}\label{ufdOV}
\eqalign{
H^{t+1}_n + \max(0,H^{t-m+1}_{n+1}-C-G) - \max(0,H^{t-m+1}_{n+1}-C) \cr
\qquad\qquad= H^t_n + \max(0,H^{t-m}_n-C-G) - \max(0,H^{t-m}_n-C),
}
\end{equation}
in the ultra-discrete limit.

\subsection{Exact solution of the ultra-discrete delayed OV model}

By taking the ultra-discrete limit, we obtain exact solutions of the ultra-discrete delayed OV model \eref{ufdOV} from the solutions \eref{sol1} and \eref{sol2} of the discrete delayed OV model.

According to the above procedure to obtain the ultra-discrete delayed OV model, we transform \eref{sol1} and \eref{sol2} into
\begin{equation}\label{sol1u}
\eqalign{
\tilde{u}^t_n
&=-\frac{L-1-4\gamma(L-L^{-m})+[L-1-4\gamma(L^{m+1}-1)] K^nL^{t-m-1}}{(1-4\gamma)(L-1)(1+K^nL^{t-m-1})}\\
&=\frac{4\gamma\frac{L-L^{-m}}{L-1} -1}{1-4\gamma}
 \frac{1+ K^{n+1} L^{t-m}}{1+K^nL^{t-m-1}},
}
\end{equation}
and
\begin{equation}\label{sol2u}
\eqalign{
\tilde{u}^t_n
&=\frac{(L-1)(1+K^nL^{t-m})}{L-1-4\gamma(L-L^{-m}) +[ L-1-4\gamma(L^{m+1}-1) ] K^nL^{t-m} }\\
&=\frac1{4\gamma\frac{L-L^{-m}}{L-1} -1}
\frac{1+K^nL^{t-m}}{1+K^{n+1}L^{t-m+1}}.
}
\end{equation}
(We exploit the dispersion relation \eref{relation3}.)
By use of the function $f(L)$ defined in \eref{f}, we have
\begin{equation}\label{ff}
4\gamma\frac{L-L^{-m}}{L-1} -1=(1-4\gamma)\frac{1+f(L)}{1-f(L)}.
\end{equation}
From the condition \eref{condition1}, we see that \eref{ff} is positive and hence the prefactors in \eref{sol1u} and \eref{sol2u} are positive.

Now we introduce the ultra-discrete parameters as $L = e^{Q/\varepsilon}$, and $K = e^{P/\varepsilon}$.
Since $L>1$, we note that $Q>0$.
The dispersion relation \eref{relation3} is transformed into
\[\eqalign{
&KL^{-m} (1 -4\gamma) +\frac{1+L+\cdots+L^{m-1}}{1+L+\cdots+L^{m}}\\
&\qquad=\frac{1-4\gamma}L+KL^{-m} \frac{1+L+\cdots+L^{m-1}}{1+L+\cdots+L^{m}},
}\]
and yields
\[
\eqalign{
&\max(P-mQ-G,\max(0,(m-1)Q)-\max(0,mQ))\\
&\qquad =\max(-G-Q,P-mQ+\max(0,(m-1)Q)-\max(0,mQ)),
}
\]
in the ultra-discrete limit.
Note that $G>0$, and one may reduce this equation to {\it the ultra-discrete dispersion relation}:
\begin{equation}
\label{udDR}
\max(Q-G,mQ-P)=0.
\end{equation}

Then let
\[
4\gamma\frac{L-L^{-m}}{L-1} -1= e^{B/\varepsilon},
\]
and this, together with \eref{relation3}, transforms into
\[KLe^{B/\varepsilon}+(1-4\gamma)(1+L+\cdots+L^m)=L+L^2+\cdots+L^{m}.\]
Thus we have
\[\max(P+Q+B, -G+\max(0,mQ) )=\max(Q, mQ)\]
in the ultra-discrete limit.
Since $G>0,\ Q>0$ and $m\in\mathbb{Z}_{>0}$, we compare both sides and consequently obtain
\[
B=-P+(m-1)Q.
\]
Finally we obtain exact solutions of the ultra-discrete delayed OV model \eref{ufdOV} by ultra-discretizing \eref{sol1u} and \eref{sol2u}:
\begin{eqnarray}
\eqalign{
&H^t_n = C +G -P +(m-1)Q + \max(0, (n+1)P+(t-m)Q )\\
&\qquad\qquad - \max(0, nP+(t-m-1)Q ),
}\label{sol6}\\
\eqalign{
&H^t_n = C + P - (m-1)Q + \max(0, nP+(t-m)Q)\\
&\qquad\qquad - \max(0, (n+1)P+(t-m+1)Q ).
}\label{sol7}
\end{eqnarray}
These are surely the solutions of \eref{ufdOV} as far as the dispersion relation \eref{udDR} holds.

The following conditions are added so that the headway $H^t_n$ should take positive values. 
The range of $L$ given in \eref{Lhat} or \eref{Lhat2} determines the range of $Q$ for \eref{sol6} or \eref{sol7} respectively in the ultra-discrete limit.
From \eref{Lhat} we have
\[
e^{2c}>\frac{1-4\gamma}{4\gamma\frac{L-L^{-m}}{L-1} -1}.
\]
Then, let $c=C/2\varepsilon$ and we have, together with the above results, the condition,
\[
C+G-P+(m-1)Q >0,
\]
for \eref{sol6} in the ultra-discrete limit.
As well, from \eref{Lhat2} we have the condition
\[
C>mQ,
\]
for \eref{sol7}.

\begin{figure}[htb]
\includegraphics[width=\textwidth]{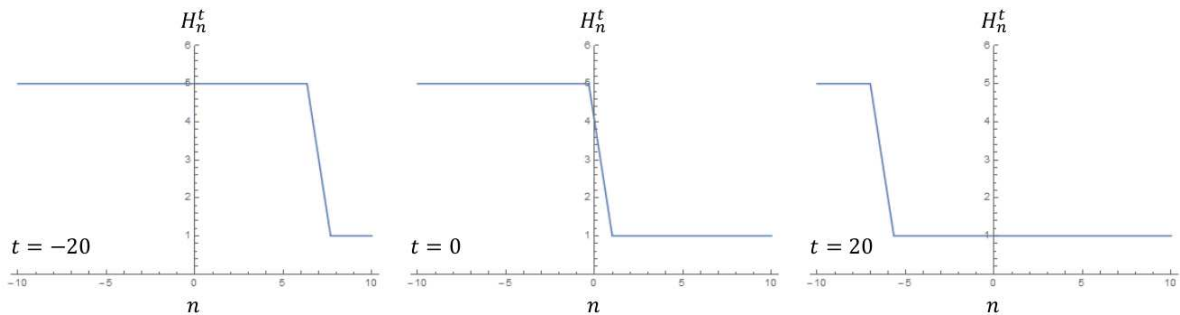}
\caption{Exact shock solution for the ultra-discrete delayed OV model: the headway $H_n^t$ \eref{sol7} of $n$th car with parameters $C=4$, $P=3$, $Q=1$, $m=3$ at time $t=-20$, $t=0$, $t=20$.}
\label{figultra}
\end{figure}

Figure \ref{figultra} shows an exact shock solution \eref{sol7}.
It provides a quite simple expression of traffic jam propagation.

\section{Summary and conclusion}

In this work, we consider the discrete delayed OV model described by a delay difference equation, where the time variable is fully discretized and the delay is hence given by an integer.
In the continuum limit, the model reduces to the delayed OV model, one of classical traffic-flow models, which is described by a delay differential equation.
We moreover work in the ultra-discrete limit where the dependent variable is also discretized to take integer values, proposing a novel traffic-flow model described by a delay cellular automaton, which we call the ultra-discrete delayed OV model.

We find exact solutions presenting a traffic jam of a shock wave form for the discrete delay model considered.
Working in the continuum and ultra-discrete limit, we also obtain the corresponding solutions to the delay differential equation and the delay cellular automaton.
To our knowledge, this is the first case of a delay cellular automaton with an exact solution.

The significance of the exact solutions presently obtained lies in that, as well as the shock formation of vehicular traffic, we have the velocity of a traffic jam propagating in an exact form from the dispersion relation.
The shock velocity implies that a quick response of drivers to the change of car distance does not always result in fast dissolution of a traffic jam.
We think that moderate response to the deceleration of cars in front will be effective to suppress disturbance of a homogeneous flow.
We note that in previous works \cite{Kanai2006b,Kanai2007a,Kanai2010}, in contrast with the present study, they give exact results for the fundamental diagram, i.e., the density-flow plot to investigate a phase transition from homogeneous to inhomogeneous density flow; but however these results do not show the detailed form of a traffic jam.

The exact solutions can also prove, as discussed in \cite{Whitham},  that nonlinear waves such as a shock and a solitary wave may appear in the range of parameters where the linear solution, which presents a uniform flow, exists stably.
This means that we can explore nonlinear waves beyond the linear stability theory.
A detailed discussion on this point will be found in our subsequent publications.

We finally add that the present work is motivated by the fact that the delayed OV model shares the same solitary-wave solution with the Kac--van Moerbeke system, one of integrable lattice systems, and we have expected that the soliton theory including discretization and ultra-discretization of integrable systems applies well to the present model.

\section*{Acknowledgments}

This work was supported in part by KAKENHI Grant Number 26610033.


\end{document}